\documentclass[aps,draft,twocolumn]{revtex4}

\input epsf

\begin{document}

\title{
Density-density propagator for one-dimensional interacting spinless fermions
with non-linear dispersion and calculation of the Coulomb drag resistivity
}
\author{A.V. Rozhkov}

\affiliation{
Institute for Theoretical and Applied Electrodynamics, ul. Izhorskaya
13/19, 124512, Moscow, Russian Federation
}

\begin{abstract}
Using bosonization-fermionization transformation we map the 
Tomonaga-Luttinger model of spinless fermions with non-linear dispersion
on the model of fermionic quasiparticles whose interaction is irrelevant in
the renormalization group sense. Such mapping allows us to set up an
expansion for the density-density propagator of the original
Tomonaga-Luttinger Hamiltonian in
orders of the (irrelevant) quasiparticle interaction. The lowest order term
in such an expansion is proportional to the propagator for free fermions.
The next term is also evaluated. The propagator found is used for
calculation of the Coulomb drug resistivity $r$ in a system of two
capacitively coupled one-dimensional conductors. It is shown that $r$ is
proportional to $T^2$ for both free and interacting fermions. The marginal
repulsive in-chain interaction acts to reduce $r$ as compared to the
non-interacting result. The correction to $r$ due to the quasiparticle
interaction is found as well. It scales as $T^4$ at low temperature.
\end{abstract}

\date{\today}

\maketitle
\hfill

\section{Introduction}

The bosonization has been an indispensable tool for one-dimensional (1D)
interacting fermion studies. The advantage of the bosonization is that it
allows to treat exactly the fermion interaction operator. This operator is
marginal in the renormalization group (RG) sense, and therefore cannot be
dealt with the help of perturbative approximations.

However the bosonization becomes inconvenient when one needs to go beyond
the marginal operators. For example, when calculating the Coulomb drag
resistivity in the system of two 1D wires
\cite{coulomb,coulomb2,derivation,coulomb3},
it is necessary to
account for the dispersion curvature $v_{\rm F}'$ of the fermions
\cite{glazman} which is an irrelevant operator. In the language of the
bosonization such operator introduces interaction between the bosons with
the coupling constant
$v_{\rm F}'$. This destroys the solubility of the bosonized Hamiltonian. 

Moreover the Coulomb drag resistivity $r$ is non-analytical in $v_{\rm F}'$:
\begin{eqnarray}
r \propto |v_{\rm F}'|.
\end{eqnarray}
This indicates that simple perturbation theory in orders of $v_{\rm F}'$ is
not applicable. 

Different methods was used to address the issue of the non-linear dispersion
\cite{glazman,glazman2,kopietz,kopietz2,samokhin,teber,teber2,pustilnik,
aristov,pereira}.
Unfortunately, these papers either rely on numerical calculation or exact
solubility or employ uncontrollable approximations or devise methods
suitable for a particular task at hand.
No universal approach emerged from those works.

A detour around the bosonization was proposed in \cite{rozhkov,rozhkov2}
where it was shown that a generic Tomonaga-Luttinger (TL) model of 1D
interacting spinless fermions may be mapped on a system of free fermionic
quasiparticles with weak irrelevant (in RG sense) interactions. 

The latter approach was particularly convenient for evaluation of the
density-density correlation function. It was demonstrated that this
correlation function is proportional to the density-density correlation
function of the free fermions plus small corrections due to the interactions
between the quasiparticles. In Ref. \cite{rozhkov,rozhkov2} the
density-density retarded propagator 
$D_{k\omega}$
and the density spectral function 
$B_{k\omega} = - 2 {\rm Im}\/\/ D_{k\omega}$
were determined to zeroth order in the quasiparticle interaction.

Since the Coulomb drag resistivity is a functional of $B_{k\omega}$, it is
natural to apply the method of \cite{rozhkov,rozhkov2} to the problem
of calculating the Coulomb drag resistivity. This is the purpose of this
paper. More specifically, we show below that at small temperature 
$r \approx a T^2$,
where the coefficient $a$ is a decreasing function of the in-chain repulsion,
and that correction due to the quasiparticle interactions $\delta r$
vanishes quicker than $T^2$: 
$\delta r \propto T^4$.
These are two main results derived below.

This paper corrects \cite{drag_PRB} and the previous version of this
preprint \cite{drag_arXiv}. In these two publications the zero-temperature
form of the spectral function is used for calculation of the
finite-temperature drag. This mistake is not crucial for it does not alter
neither $T^2$ behavior of $r$, nor $T^4$ behavior of $\delta r$. It does
change the $O(1)$ numerical coefficient in front of formula for $r$, as well
as details of $r$ and $\delta r$ derivations.

The presentation is structured as follows. First in
Sect. \ref{mapping} we establish the mapping of the TL model on the
quasiparticle model. Next in Sect. \ref{propagator} we obtain the
density-density propagator and the Coulomb drag resistivity to zeroth
and first order in the quasiparticle interaction. The conclusions of the
paper are given in Sect. \ref{conclusions}. Involved calculations are
performed in Appendices.

\section{The mapping}
\label{mapping}

In this paper we study the model of 1D fermions whose Hamiltonian is:
\begin{eqnarray}
H &=& H_{\rm kin} + H_{\rm nl} + H_{\rm int},
\label{H}
\\
H_{\rm kin}
&=&
{\rm i} v_{\rm F}
\int dx
\left(
        \colon
        \psi^\dagger_{{\rm L}}
        \nabla\psi^{\vphantom{\dagger}}_{{\rm L}}
        \colon
        -
        \colon
        \psi^\dagger_{{\rm R}}
        \nabla\psi^{\vphantom{\dagger}}_{{\rm R}}
        \colon
\right),
\label{Hkin}
\\
H_{\rm nl}
&=&
v_{\rm F}'
\int dx
\left(
        \colon
        (\nabla \psi^\dagger_{{\rm L}})
        (\nabla\psi^{\vphantom{\dagger}}_{{\rm L}})
        \colon
        +
        \colon
        (\nabla \psi^\dagger_{{\rm R}})
        (\nabla \psi^{\vphantom{\dagger}}_{{\rm R}})
        \colon
\right),
\quad
\label{Hnl}
\\
H_{\rm int}
&=&
g \int dx
\rho_{\rm R}
\rho_{\rm L},
\end{eqnarray} 
where 
$\psi_{\rm R,L}$
are chiral fermionic fields corresponding to the right-moving (subscript `R')
and left-moving (subscript `L') fermions, 
$\rho_{\rm R,L} = \colon \psi^\dagger_{\rm R,L}
\psi^{\vphantom{\dagger}}_{\rm R,L} \colon$
are chiral fermion densities, and the colons denote normal ordering. The
cutoff $\Lambda$ is assumed for this quantum field theory.

A non-perturbative approach to handle Hamiltonian $H$ was proposed in
Ref. \cite{rozhkov}. There a unitary operator $U$ was constructed which
transforms $H$ into the quasiparticle Hamiltonian:
\begin{eqnarray}
H_{\rm qp}
=
\tilde H_{\rm kin}
+
\tilde H_{\rm nl}
+
\tilde H_{\rm int}'
+
\delta \mu (N_{\rm R} + N_{\rm L}),
\label{Hqp}
\end{eqnarray}
where $N_{\rm R, L}$ are the total number of right-moving (left-moving)
fermions, $\tilde H_{\rm kin}$ and $\tilde H_{\rm nl}$ have the same form as
$H_{\rm kin}$ and $H_{\rm nl}$ but with $\tilde v_{\rm F}$ and $\tilde
v_{\rm F}'$ instead of $v_{\rm F}$ and $v_{\rm F}'$:
\begin{eqnarray}
\tilde H_{\rm kin} 
&=&
{\rm i} \tilde v_{\rm F}
\int dx
\left(
        \colon
        \tilde \psi^\dagger_{{\rm L}}
        \nabla \tilde \psi^{\vphantom{\dagger}}_{{\rm L}}
        \colon
        -
        \colon
        \tilde \psi^\dagger_{{\rm R}}
        \nabla \tilde \psi^{\vphantom{\dagger}}_{{\rm R}}
        \colon
\right),
\label{Hkin_qpe}
\\
\tilde H_{\rm nl}
&=&
\tilde v_{\rm F}'
\int dx
\left(
        \colon
        (\nabla \tilde \psi^\dagger_{{\rm L}})
        (\nabla \tilde \psi^{\vphantom{\dagger}}_{{\rm L}})
        \colon
        +
        \colon
        (\nabla \tilde \psi^\dagger_{{\rm R}})
        (\nabla \tilde \psi^{\vphantom{\dagger}}_{{\rm R}})
        \colon
\right),
\quad
\label{Hnl_qpe}
\\
\tilde v_{\rm F} 
&=&
v_{\rm F}
\sqrt{1-\left(\frac{g}{2\pi v_{\rm F}}\right)^2},
\label{vF_ren}
\\
\tilde v_{\rm F}'
&=&
\frac{ v_{\rm F}' } { 4 }
\left(
        {\cal K}^{3/2} + 3 {\cal K}^{-1/2}
\right).
\label{vF'_ren}
\end{eqnarray}
The usual TL parameter ${\cal K}$ is used in the last formula:
${\cal K} = \sqrt{(2\pi v_{\rm F} - g)/(2\pi v_{\rm F} + g)}$.
Operators with tildes correspond to the quasiparticles: $\tilde \psi_p$ is
the quasiparticle field, below we use $\tilde \rho_p$, which is the
quasiparticle density.

The quasiparticle interaction $\tilde H_{\rm int}'$ in Eq.(\ref{Hqp}) is
given by expression:
\begin{eqnarray}
\tilde H_{\rm int}'
=
- \sum_p i p \tilde g' 
\int dx \tilde \rho_{-p} 
\left[ 
	\colon 
		\tilde \psi^\dagger_p
		(\nabla \tilde \psi^{\vphantom{\dagger}}_p) 
	\colon 
	-
	\colon 
		(\nabla \tilde \psi^\dagger_p) 
		\tilde \psi^{\vphantom{\dagger}}_p
	\colon 
\right],
\label{Hint'}
\\
\tilde g' 
=
\frac{
	\pi v_{\rm F}'
     }
     {
	2
     }
\left(
	{\cal K}^{3/2} - {\cal K}^{-1/2}
\right).
\end{eqnarray}
In Eq.(\ref{Hint'}) the summation runs over the chirality index $p={\rm
R,L}$ whose numerical values are $p=+1$ for `R' and $p= -1$ for `L'.
Observe that the operator 
$\tilde H'_{\rm int}$
is irrelevant: its scaling dimension is equal to 3 which is greater than 2
\cite{boson}.
(To evaluate the scaling dimension one has to add together the scaling
dimension of two fermion operators, the gradient, and the electron
density operator: $1/2 + 1/2 + 1 + 1 = 3$).

The chemical potential shift $\delta \mu$ in Eq.(\ref{Hqp}) is also
induced by the transformation $U$. However since such shift causes nothing
but additional renormalization of the quasiparticle dispersion parameters
$\tilde v_{\rm F}$ and $\tilde v_{\rm F}'$, we do not keep track of it
below.

The readers who are interested to learn how operator $U$ is constructed
should consult Ref.\cite{rozhkov,rozhkov2}. Here we use
bosonization-fermionization sequence to establish the desired equivalence
between $H$ and $H_{\rm qp}$. That way, we derive the result we need with
no reference to the new technique unfamiliar to the majority of the
researchers in the field. 

The bosonization prescription \cite{boson,giamarchi} expresses the 1D chiral
fermion field as an exponential of the Bose field $\Phi$ and its conjugate
$\Theta$:
\begin{eqnarray}
\psi^\dagger_{p} (x) =
(2\pi a)^{-1/2} \eta_{p}{\rm e}^{{\rm i}\sqrt{\pi} \left[
\Theta(x) + p\Phi(x) \right]},\label{bos}
\end{eqnarray}
where $a \propto 1/\Lambda$ and $\eta_p$ is the Klein factor. Consequences
of this formula are:
\begin{eqnarray}
\rho_p 
= 
\frac{1}{2\sqrt{\pi}}
\left(
	\nabla \Phi
	+
	p \nabla \Theta
\right),
\label{rho_bos}
\\
-ip
\colon 
	\psi^{{\dagger}}_p
	(\nabla \psi^{\vphantom{\dagger}}_p)
\colon
+
\text{H.c.}
=
\frac{1}{2}
\colon
	\left(
		\nabla \Phi
		+
		p \nabla \Theta
	\right)^2
\colon,
\label{kin_bos}
\\
\colon 
	(\nabla \psi^{{\dagger}}_p)
	(\nabla \psi^{\vphantom{\dagger}}_p)
\colon
-
\frac{1}{6}
\nabla^2 \rho_p
=
\frac{\sqrt{\pi}}{6}
\colon
	\left(
		\nabla \Phi
		+
		p \nabla \Theta
	\right)^3
\colon.
\label{nl_bos}
\end{eqnarray}
Using the above formulas we can write the bosonic form of $H$:
\begin{eqnarray}
H 
\left[ 
	\Phi, \Theta
\right] 
=  
H_{\rm kin} 
\left[ 
	\Phi, \Theta
\right] 
+ 
H_{\rm int} 
\left[ 
	\Phi, \Theta
\right] 
+ 
H_{\rm nl}
\left[ 
	\Phi, \Theta
\right] , 
\label{Hbos}
\\
H_{\rm kin} + H_{\rm int} 
=  
\frac{\tilde v_{\rm F}}{2} 
\int dx 
\left[
	{\cal K} 
	\colon
		\left( \nabla \Theta \right)^2 
	\colon
	+
	{\cal K}^{-1} 
	\colon
		\left( \nabla \Phi \right)^2 
	\colon
\right], 
\label{Htl_bos}
\\
H_{\rm nl} 
=  
\frac{ \sqrt{\pi} } { 6 }
v_{\rm F}' \int dx
\sum_p
	\colon
		\left( 
			\nabla \Phi + p \nabla \Theta
		\right)^3
	\colon.
\label{Hnl_bos}
\end{eqnarray}
Eq.(\ref{Hnl_bos}) was derived in Ref. \cite{haldane}.

Once the bosonic form is explicitly written we are ready for the second
step of the derivation -- rescaling of the bosonic fields: 
$\tilde \Phi = {\cal K}^{-1/2} \Phi$,
$\tilde \Theta = {\cal K}^{1/2} \Theta$.
The different pieces of Eq. (\ref{Hbos}) can be expressed in terms of this
new boson as such:
\begin{eqnarray}
H_{\rm kin} + H_{\rm int} 
=  
\frac{\tilde v_{\rm F}}{2} 
\int dx 
\left[
	\colon
		\left( \nabla \tilde \Theta \right)^2 
	\colon
	+
	\colon
		\left( \nabla \tilde \Phi \right)^2 
	\colon
\right], \quad
\\
H_{\rm nl} 
=  
\frac{ \sqrt{\pi} } { 6 }
v_{\rm F}' \int dx
\sum_p
	\colon
		\left( 
			{\cal K}^{1/2} \nabla \tilde \Phi
			+
		 	p {\cal K}^{-1/2}\nabla \tilde \Theta 
		\right)^3
	\colon.
\label{Hnl_bos_tilde}
\end{eqnarray} 
Introducing a new fermion (the quasiparticle) with the help of the formula:
\begin{eqnarray}
\tilde \psi^\dagger_{p} (x) =
(2\pi a)^{-1/2} \eta_{p}{\rm e}^{{\rm i}\sqrt{\pi} \left[
\tilde \Theta(x) + p \tilde \Phi(x) \right]},
\label{quasiparticle}
\end{eqnarray}
one can refermionize $H$. Namely, inverting Eq.(\ref{kin_bos}), we obtain
for the sum 
$H_{\rm kin} + H_{\rm int}$:
\begin{eqnarray}
H_{\rm kin} + H_{\rm int} = \tilde H_{\rm kin},
\end{eqnarray}
where 
$\tilde H_{\rm kin}$
is given by Eq.(\ref{Hkin_qpe}).
On the right-hand side of this expression the interaction term 
$\int dx \tilde \rho_{\rm R} \tilde \rho_{\rm L}$
is absent. Thus, the marginal interaction, the most troublesome part of the
Hamiltonian, is removed. 

The price we have to pay for the absence of the marginal interaction is that
$H_{\rm nl}$
expressed in
$\tilde \Phi$, $\tilde \Theta$,
Eq.(\ref{Hnl_bos_tilde}), cannot be easily fermionized. It is convenient
to rewrite the latter equation:
\begin{eqnarray}
H_{\rm nl}
&=&
\frac{ \sqrt{\pi} } { 6 }
\tilde v_{\rm F}' \int dx
\sum_p
	\colon
		\left( 
			\nabla \tilde \Phi + p \nabla \tilde \Theta
		\right)^3
	\colon
\label{Hnl_bos_aux}
\\
\nonumber
&+&
\frac{\tilde g'}{4\sqrt{\pi}}
\int dx
\sum_p
	\colon
		\left(
			\nabla \tilde \Phi 
			+
			p \nabla \tilde \Theta
		\right)^2
		\left(
			\nabla \tilde \Phi
			-
			p \nabla \tilde \Theta
		\right)
	\colon.
\end{eqnarray}
The advantage of this form is that it depends on 
$( \nabla \tilde \Phi \pm \nabla \tilde \Theta)$ combinations only. Thus
Eqs.(\ref{rho_bos}), (\ref{kin_bos}) and (\ref{nl_bos}) may be immediately
applied and $H_{\rm nl}$ may be fermionized:
\begin{eqnarray}
H_{\rm nl} = \tilde H_{\rm nl} + \tilde H_{\rm int}',
\end{eqnarray}
where 
$\tilde H_{\rm nl}$ 
and
$\tilde H_{\rm int}'$
are given by Eqs. (\ref{Hnl_qpe}) and (\ref{Hint'}). 

This almost concludes the derivation of Eq.(\ref{Hqp}). What we lack is the
$\delta \mu$ term of $H_{\rm qp}$. To obtain this term we must {\it (i)}
handle normal ordered expressions more accurately and {\it (ii)} take
special care about the zero modes $N_{\rm R,L}$. Technically this is
similar to the treatment of Ref. \cite{rozhkov,rozhkov2}. Since we are not
interested in the chemical potential shift, we do not address this issue
here.

\section{Density-density propagator and the Coulomb drag resistivity}
\label{propagator}

Once the mapping of $H$ on $H_{\rm qp}$ is established, we can use it to
calculate the density-density propagator 
$D_{k\omega}$. 
First let us find the following density-density correlation function:
\begin{eqnarray}
{\cal R} = 
\langle
	[\rho_{\rm R} (x,\tau) 
	+
	\rho_{\rm L} (x,\tau) ]
	[\rho_{\rm R} (0,0)
	+
	\rho_{\rm L} (0,0) ]
\rangle.
\end{eqnarray}
In the bosonic form it equals to:
\begin{eqnarray}
{\cal R} 
= 
\frac{1}{\pi}
\langle
	\nabla \Phi (x)
	e^{-\tau H}
	\nabla \Phi (0)
\rangle.
\end{eqnarray}
After the field rescaling it becomes:
\begin{eqnarray}
{\cal R} 
= 
\frac{\cal K}{\pi}
\langle
	\nabla \tilde \Phi (x)
	e^{-\tau H_{\rm qp}}
	\nabla \tilde \Phi (0)
\rangle.
\end{eqnarray}
Under fermionization ${\cal R}$ transforms into:
\begin{eqnarray}
{\cal R} = 
{\cal K}
\langle
	[\tilde \rho_{\rm R} (x,\tau) 
	+
	\tilde \rho_{\rm L} (x,\tau) ]
	[\tilde \rho_{\rm R} (0,0)
	+
	\tilde \rho_{\rm L} (0,0) ]
\rangle_{\rm qp},
\end{eqnarray}
where the subscript `qp' reminds that the averaging is to be performed with
respect to the quasiparticle Hamiltonian $H_{\rm qp}$.

One can prove through the same line of reasoning that 
\begin{eqnarray}
D_{k\omega} = {\cal K}
\tilde D_{k\omega},
\end{eqnarray}
where 
$\tilde D_{k\omega}$
is the retarded density-density propagator for the quasiparticle Hamiltonian 
$H_{\rm qp}$.
Consequently, the task of finding the density-density propagator for the
physical fermions is reduced to the task of finding the quasiparticle
density-density
propagator. The latter is much easier, for the quasiparticle interaction 
$\tilde H_{\rm int}'$
is irrelevant in the RG sense.

As a starting point we calculate 
$D_{k\omega}$
to zeroth order in 
$\tilde H_{\rm int}'$: 
\begin{eqnarray}
D_{k \omega} = {\cal K}\left(
				\tilde D_{{\rm R} k \omega}^0 +
				\tilde D_{{\rm L} k \omega}^0 
			\right) + O(\tilde H'_{\rm int}),
\label{Dret}
\\
B_{k \omega} = {\cal K}\left(
				\tilde B_{{\rm R} k \omega}^0 +
				\tilde B_{{\rm L} k \omega}^0 
			\right) + O(\tilde H'_{\rm int}),
\label{B}
\end{eqnarray}
where the chiral quasiparticle density-density propagators 
$\tilde D_{\rm R,L}^0$ and corresponding spectral densities
$\tilde B_{\rm R,L}^0$ are:
\begin{eqnarray}
\tilde D^0_{pk\omega} =
\int_{q} \frac{
		n_{pq-k/2} - n_{pq+k/2} 
	      }
	      {
		\omega + \tilde \varepsilon_{pq-k/2} 
		- \tilde \varepsilon_{pq+k/2} + i0
	      },
\label{D_qpp}
\\
\tilde B^0_{pk\omega} =
\int_{-\Lambda}^\Lambda dq
\left(
	n_{pq-k/2} - n_{pq+k/2} 
\right) 
\times
\label{B_qpp}
\\
\nonumber 
	 \delta( 
		\omega + \tilde  \varepsilon_{pq-k/2}
		- \tilde \varepsilon_{pq+k/2} 
	       ).
\end{eqnarray}
Symbol
$\int_q \ldots$
stands for 
$(2\pi)^{-1} \int_{- \Lambda}^{\Lambda} dq \ldots$,
quantity $\tilde \varepsilon_{pq}$
is the quasiparticle dispersion, and $n_{pq}$ is the Fermi occupation number:
\begin{eqnarray}
\tilde \varepsilon_{pq} = p \tilde v_{\rm F} q + \tilde v_{\rm F}' q^2,
\label{dispersion}
\\
n_{pq} = \left[
		1 + \exp(\tilde \varepsilon_{pq}/T)
      \right]^{-1}.
\end{eqnarray} 
At $T=0$ the integrals in Eq.~(\ref{D_qpp}) and Eq.~(\ref{B_qpp}) can be
easily evaluated. One finds the lowest order expression for $D$ and $B$
\cite{rozhkov,rozhkov2}:
\begin{eqnarray}
D_{k\omega}
= 
\frac{{\cal K}} {4\pi \tilde v'_{\rm F} k}
\ln
\left[ 
        \frac{ 
                ( \tilde v_{\rm F} k - \tilde v'_{\rm F} k^2)^2 
                - 
                (\omega + i0)^2
             }
             { 
                (\tilde v_{\rm F} k + \tilde v'_{\rm F} k^2)^2 
                - 
                (\omega +i0)^2
             }
\right]
\label{DT=0}
\\
\nonumber
+O(\tilde H'_{\rm int}),
\\
B_{k\omega} 
= 
\frac{{\cal K}} {2 \tilde v'_{\rm F} k} 
\left[
        \vartheta \left( 
                        \omega^2 
                        - 
                        (\tilde v_{\rm F} k - \tilde v'_{\rm F} k^2)^2
                  \right) 
\right.
\\
\nonumber
\left.
        -
        \vartheta \left( 
                        \omega^2 
                        - 
                        (\tilde v_{\rm F} k + \tilde v'_{\rm F} k^2)^2
                  \right)
\right] 
{\rm sgn\ } \omega
+O(\tilde H'_{\rm int}).
\end{eqnarray}

For non-zero temperature Aristov shows \cite{aristov} that Eq.~(\ref{D_qpp})
reduces to an expression with a special function. Eq.~(\ref{B_qpp}) may be
calculated even for non-zero temperature in terms of elementary functions:
\begin{eqnarray} 
\label{B_finT}
\tilde B_{pk\omega}^0
=
\frac{
	\sinh\left(
		\frac{\omega}{2T}
	     \right)
     }
     {
	4 |\tilde v_{\rm F}' k |
	\cosh(\tilde v_{\rm F} 
	      \frac{
			\delta \omega_p - \tilde v_{\rm F}' k^2
		   }
		   {
			4 \tilde v_{\rm F}' k T
		   }
	     )
	\cosh(\tilde v_{\rm F}
	      \frac{
			\delta \omega_p + \tilde v_{\rm F}' k^2
		   }
		   {
			4 \tilde v_{\rm F}' k T
		   }
	     )
     },
\\
\delta \omega_p = \omega - p \tilde v_{\rm F} k.
\end{eqnarray}
Appendix~\ref{app_B0} provides the details of the calculation.
Eq.~(\ref{B_finT}) can be used to find $B_{k\omega}$, Eq.~(\ref{B}).

Once the spectral function is found the Coulomb drag resistivity may be
evaluated with its help. Before proceeding with such calculations, let us
briefly explain what Coulomb drag is.

In the Coulomb drag experiment two parallel 1D wires (subscript $i=1,2$) of
length $L$ with
Hamiltonians $H$, Eq.(\ref{H}), are coupled capacitively with the Hamiltonian 
$H_{\rm C} = g_{\rm C} \int dx \rho_1 \rho_2$. 
Because of this coupling, electrical current $I$ in one of the wires induces
potential drop $V$ across the other wire. The proportionality coefficient
between $V$ and $I$ is called the Coulomb drag resistivity 
$r = V/IL$. 
It characterizes the pulling force, which the fermions in the current-carrying
wire exert on the fermions in the other wire. 

The experimental physics of the Coulomb drag is quite rich: the observed
values of $r$ could be either positive or negative, and show dependence on
temperature, spacial inhomogeneity, and applied magnetic field.

In general, two mechanisms are discussed in the theoretical literature
\cite{derivation}.
According to one mechanism, the drag occurs because Wigner crystal-like
correlations in both wires lock against each other. In the RG language this
corresponds to the relevance of the inter-wire backscattering interaction.
Such mechanism works best at zero temperature but quickly deteriorates at
$T>0$. We do not study this mechanism in our paper.

The second mechanism is insensitive to the inter-wire backscattering.
Instead it relies upon interaction of the smooth components of the electron
densities. It is more resilient towards temperature but the resultant value
of $r$ is proportional to 
$| v_{\rm F}' |$.
Because of this, it is impossible to study the second mechanism within the
well-established framework of the one-dimensional bosonization. The purpose
of this paper is to provide a reliable approach overcoming this difficulty.

We start with the following formula for $r$ (Eq.(7) of Ref. \cite{glazman})
which is valid when the inter-wire backscattering can be neglected:
\begin{eqnarray}
r=
\frac{
	g_{\rm C}^2 
     }
     {
	16 \pi^3 n^2  T
     }
\int_0^{+\infty} dk \int_0^{+\infty} d\omega
\frac{
	k^2 B_{1,k\omega} B_{2,k\omega}
     }
     {
	{\rm sh}^2 (\omega/2T)
     }.
\label{r_def}
\end{eqnarray}
Here $n$ is the electron density.
The spectral density $A (k, \omega)$ of Ref. \cite{glazman} relates to our
spectral density $B_{k\omega}$ as:
$B_{k\omega} = -2 A(k, \omega)$.
In that reference the notation $U_{12}$ is used for the inter-wire coupling
constant $g_{\rm C}$. Below we study the case of identical wires:
$B_1 = B_2 = B$. 
We assume that $|\tilde v_{\rm F}'|$ is small:
\begin{eqnarray}
|\tilde v_{\rm F}'| \Lambda < \tilde v_{\rm F}.
\end{eqnarray} 

Since we know 
$B_{k\omega}$
to zeroth order in $\tilde g'$, it is straightforward to find $r$ with the
same accuracy. Let us begin our calculation with the following observation.
The function:
\begin{eqnarray}
\tilde b_{pk\omega}
=
\frac{
	\tilde B^0_{pk\omega} 
     }
     {
	\sinh\left(
		\frac{\omega}{2T}
	     \right)
     }
\\
\nonumber
=
\frac{
	1
     }
     {
	2 |\tilde v_{\rm F}' k |
	\left[
		\cosh(\frac{
				\tilde v_{\rm F} \delta \omega_p 
			   }
		   	{
				2 \tilde v_{\rm F}' k T
		   	}
	     	     )
		+
		\cosh(\frac{
				\tilde v_{\rm F} k
		   	   }
		   	   {
				2T 
		   	   }
	     )
	\right]
     }
\label{B_finT2}
\end{eqnarray} 
localized mostly at small momenta and near the `light cone':
\begin{eqnarray}
|k|<T/\tilde v_{\rm F},
\\
|\omega - p \tilde v_{\rm F} k| < \tilde v_{\rm F}' k T /\tilde v_{\rm F} 
< \tilde v_{\rm F}' T^2 / \tilde v_{\rm F}^2.
\label{region_kw}
\end{eqnarray}
Outside of this region 
$\tilde b_{pk\omega}$
is exponentially small.

Since 
$\tilde b_{{\rm L} k\omega}$
is exponentially small when $k$ and $\omega$ are positive, for the purpose of
calculating the integral Eq.~(\ref{r_def}) we may further approximate: 
$B \approx {\cal K} \tilde B_{\rm R}^0$. 
Substituting this in Eq.~(\ref{r_def}) we obtain:
\begin{eqnarray}
r
=
\frac{
	g_{\rm C}^2 {\cal K}^2
     }
     {
	256 \pi^3 (\tilde v_{\rm F}')^2 n^2 T
     }
I(T),
\\
I(T)
=
16 (\tilde v_{\rm F}')^2 
\int_0^{+\infty} dk
\int_{0}^{+\infty} d\omega
k^2 \tilde b^2_{{\rm R} k\omega}
\label{r^0}
\\
\nonumber 
=
\int
   \frac{dk d\omega}
	{
		\cosh^{2}(\tilde v_{\rm F} 
			      \frac{
				     \omega - \tilde v_{\rm F} k 
					- \tilde v_{\rm F}' k^2
		   		   }
		   		   {
					4 \tilde v_{\rm F}' k T
				   }
	     		     )
		\cosh^{2}(\tilde v_{\rm F}
	     		      \frac{
				     \omega - \tilde v_{\rm F} k 
					+ \tilde v_{\rm F}' k^2
		                   }
		                   {
					4 \tilde v_{\rm F}' k T
		                   }
	                     )
	}.
\end{eqnarray}
The integral $I(T)$ is evaluated in Appendix~\ref{app_I(T)}. It is shown
that, if temperature is low:
$T \ll \tilde \epsilon_F$,
where renormalized Fermi energy is:
\begin{eqnarray} 
\tilde \epsilon_{\rm F}
=
\frac{\tilde v_{\rm F}^2}{ 4 \tilde v_{\rm F}'},
\end{eqnarray} 
then
\begin{eqnarray}
I = \frac{
		32 |\tilde v_{\rm F}'| T^3
	 }
	 {
		\tilde v_{\rm F}^3
	 } + o(T^3).
\end{eqnarray}
This gives us:
\begin{eqnarray}
r = \frac{
		g_{\rm C}^2 {\cal K}^2 
	 }
	 {
		8 \pi^3 \tilde v_{\rm F}^3 | \tilde v_{\rm F}'| n^2
	 }
T^2 + o(T^2).
\label{r_free}
\end{eqnarray} 
The electron zero-temperature density, which enters Eq. (\ref{r_free}), is
not an independent quantity. It can be expressed in terms of the dispersion
parameters: 
$n = v_{\rm F}/(2 \pi |v_{\rm F}'|)$.

Furthermore, it is possible to show with the help of Eqs. (\ref{vF_ren}) and
(\ref{vF'_ren}) that
$v_{\rm F} = \tilde v_{\rm F} + O(g^2)$ 
and
$v_{\rm F}' = \tilde v_{\rm F}' + O(g^2)$.
Since our accuracy does not allow us to keep $O(g^2)$ terms, we can replace
$\tilde v_{\rm F}$ by $v_{\rm F}$ and $\tilde v_{\rm F}'$ by $v_{\rm F}'$
in Eq.(\ref{r_free}). Thus, it is true:
\begin{eqnarray}
r & \approx & a T^2,
\label{r_free_approx}
\\
a &=&
    \frac{
		g_{\rm C}^2 {\cal K}^2 | v_{\rm F}'|
	 }
	 {
		2 \pi v_{\rm F}^5
	 }
    + O(g^2).
\label{a}
\end{eqnarray}
The interaction enters the expression for $a$ through 
${\cal K}^2 = 1 -  g / (\pi v_{\rm F}) + O(g^2)$.
We see that the repulsive in-chain interaction ($g>0$) acts to reduce $r$.

Eq.(\ref{r_free_approx}) may be cast in the following form:
\begin{eqnarray}
r \approx \frac{c_1 {\cal K}^2}{l_0}
    \left(
		\frac{T}{\epsilon_{\rm F}}
    \right)^2,
\label{r_glazman}
\end{eqnarray}
where $l_0^{-1}$ and $\epsilon_{\rm F}$ are defined in Ref. \cite{glazman}.
They are:
\begin{eqnarray}
\epsilon_{\rm F} = \tilde \epsilon_{\rm F} + O(g^2),
\\
\frac{1}{l_0}
=
\left(
	\frac{g_{\rm C}}{2 \pi v_{\rm F}}
\right)^2
n
=
\frac{
	g_{\rm C}^2
     }
     {
	8 \pi^3 |v_{\rm F}'| v_{\rm F}^{\vphantom{'}} 
     } .
\end{eqnarray}
The numerical coefficient $c_1$ is equal to $\pi^2/4$. For free fermions 
(${\cal K} = 1$)
the formula identical to Eq. (\ref{r_glazman}), with the same value for
$c_1$ is derived in \cite{glazman}.

Furthermore, that paper establishes that 
$r \propto T^2$ 
for both non-interacting fermions and for exactly soluble
Calogero-Sutherland model. Our Eq.(\ref{r_free}) proves that $T^2$
dependence holds for a generic interacting model as well. 

In Refs.~\cite{drag_PRB,drag_arXiv}, where, mistakenly, the
zero-temperature form of $B$ is used in the integral for $r$, different
value of the coefficient $c_1$ is found. Present derivation corrects that
error.

Eq.(\ref{r_free_approx}) and Eq.~(\ref{a}) account for effects of marginal
interaction $g$. In addition to that, our method allows us to evaluate the
correction $\delta r$ to $r$ due to the quasiparticle interaction
$\tilde g'$.

To find $\delta r$ we calculate the lowest order correction to
$D^0_{k\omega}$. Such correction appears in the first order in $\tilde g'$:
since 
$\tilde H_{\rm int}'$
couples the right-moving and the left-moving quasiparticles, the expectation
value 
$\langle \tilde \rho_{\rm R} (x,\tau) \tilde \rho_{\rm L} (0,0) \rangle$
taken with respect to $H_{\rm qp}$, Eq.(\ref{Hqp}), is no longer zero but
instead
$O(\tilde g')$.
Thus, full Matsubara propagator ${\cal D} = {\cal D}^0 + \delta {\cal D}$,
where:
\begin{eqnarray}
\delta {\cal D} (x, \tau)
&=&
{\cal K} \delta \tilde {\cal D} (x, \tau)
\\
\nonumber 
& \approx &
{\cal K}
\int d\tau'
\left[
	\langle	\langle
		\tilde \rho_{\rm R} (x,\tau)
		\tilde H_{\rm int}' (\tau') 
		\tilde \rho_{\rm L} (0,0)
	\rangle \rangle_0
\right.
\label{Matsubara_corr}
\\
\nonumber
&+& 
\left.
	\langle\langle
		\tilde \rho_{\rm L} (x,\tau)
		\tilde H_{\rm int}' (\tau') 
		\tilde \rho_{\rm R} (0,0)
	\rangle \rangle_0
\right].
\end{eqnarray}
Here the symbol 
$\langle \langle \ldots \rangle \rangle_0$ 
stands for time-ordered averaging with respect to the non-interacting 
($\tilde g' = 0$)
quasiparticle Hamiltonian. Matsubara propagators are denoted by calligraphic
letters (e.g., ${\cal D}$, ${\cal P}$), retarded propagators are denoted by
italic letters (e.g., $D$, $P$).

\begin{figure} [!t]
\centering
\leavevmode
\epsfxsize=5cm
\epsfysize=5cm
\epsfbox[88 444 362 718] {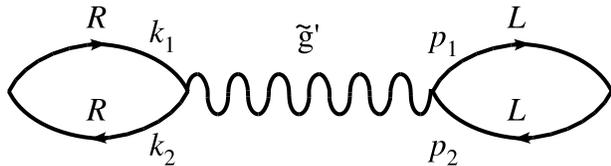}
\caption[]
{\label{diag} 
Feynman diagram corresponding to the lowest order interaction correction
to the density-density propagator.
}
\end{figure}

Feynman diagram, which describes $\delta \tilde {\cal D}$, is shown on
Fig.\ref{diag}.  The wavy interaction line is to be identified with 
$\tilde g' (k_{\rm  1} + k_{\rm  2} - p_{\rm  1} - p_{\rm  2} )$,
$k_{i}$,
$p_{i}$
are the fermion momenta.

When evaluating 
$\delta {\cal D}$ 
we note that in the non-interacting quasiparticle Hamiltonian the left-moving
and the right-moving quasiparticles are decoupled from each other.
Consequently, the object
$\langle \langle \rho_{\rm L,R} H_{\rm int}' \rho_{\rm L,R} \rangle
\rangle_0$ is split
into products of left-only and right-only expectation values. 

As a result of this derivation one can show that the lowest order correction
to the Matsubara propagator due to the quasiparticle interaction is:
\begin{eqnarray}
\delta {\cal D}_{k \omega} 
=
- 2 \tilde g'
{\cal K}
\sum_p
	\tilde {\cal D}_{p k \omega}^0 
	\tilde {\cal P}_{-p k \omega} ,
\label{deltaD}
\end{eqnarray}  
where
$\tilde {\cal D}^0_{pk\omega}$
is the chiral non-interacting Matsubara propagator, the propagator
$\tilde {\cal P}_p$
is defined as:
\begin{eqnarray}
&&\tilde {\cal P}_{p} (x , \tau )
=
- i p 
\langle \langle
		\tilde \rho_{p} (x, \tau)
\\
\nonumber
&& 
\quad
\times 
		\left\{
			\colon
				\tilde \psi_{p}^\dagger (0,0)
				[
					\nabla \tilde 
					\psi^{\vphantom{\dagger}}_p (0,0)
				]
			\colon
			-
			\colon
				[ \nabla \tilde \psi_{p}^\dagger (0,0) ]
				\tilde \psi^{\vphantom{\dagger}}_p (0,0)
			\colon
		\right\}
\rangle \rangle_0 .
\end{eqnarray} 
It equals to:
\begin{eqnarray}
\tilde {\cal P}_{p k \omega}
=
- \frac{1}{2\pi \tilde v_{\rm F}'}
+ \frac{( i\omega - p \tilde v_{\rm F} k )}{p \tilde v_{\rm F}' k}
\tilde {\cal D}_{p k \omega}^0.
\end{eqnarray}

To find the correction to the spectral function $\delta B$ it is necessary
to perform the analytic continuation in Eq.(\ref{deltaD}):
$
\delta {D}_{k \omega} 
=
-2 \tilde g'
{\cal K}
\sum_p
	\tilde {D}_{-p k \omega} ^0
	\tilde {P}_{pk \omega} ,
$
where the retarded quantity 
$\tilde P_{pk\omega} = \tilde {\cal P}_{pk \omega} |_{i\omega \rightarrow
\omega + i0}$ 
is equal to:
\begin{eqnarray}
\tilde P_{pk\omega} 
=
-\frac{1}{2 \pi \tilde v_{\rm F}'}
+
\frac{(\omega - p \tilde v_{\rm F} k + i0)}{p \tilde v_{\rm F}' k}
\tilde D^0_{pk\omega}.
\end{eqnarray}
We obtain for $\delta B$:
\begin{eqnarray}
\delta B_{k\omega}
=
\frac{\tilde g' {\cal K}}{\pi \tilde v_{\rm F}'}
\sum_p
	\tilde B_{-p k \omega}^0
	\left(
		4 \pi \tilde v_{\rm F} {\rm Re}\/ \tilde D_{p k\omega}^0
		+
		1
	\right).
\end{eqnarray}
This is the lowest order correction to the spectral density due to the
quasiparticle interactions.

The square of the spectral density, which enters the equation for the
Coulomb drag, is:
\begin{eqnarray}
B^2_{k\omega} \approx ({\cal K} \tilde B^0_{k\omega})^2 
\\
\nonumber 
+
\frac{2\tilde g' {\cal K}^2}{\pi \tilde v_{\rm F}'}
\sum_{p,p'}
	\tilde B_{p' k \omega}^0 \tilde B_{-p k \omega}^0
	\left(
		4 \pi \tilde v_{\rm F} {\rm Re}\/ \tilde D_{p k\omega}^0
		+
		1
	\right).
\end{eqnarray} 
We already pointed out the fact that 
$\tilde b_{{\rm L}k\omega}$
is exponentially small for positive $k$ and $\omega$. Therefore, in the
region of the integration we may neglect the terms proportional to
$\tilde B_{\rm L}^0 \tilde B_{\rm R}^0$
and
$(\tilde B_{\rm L}^0 )^2$:
\begin{eqnarray}
B^2_{k\omega} \approx ({\cal K} \tilde B^0_{{\rm R} k\omega})^2 
+
\frac{2\tilde g' {\cal K}^2}{\pi \tilde v_{\rm F}'}
	(\tilde B_{{\rm R} k \omega}^0)^2
	\left(
		4 \pi \tilde v_{\rm F} {\rm Re}\/ \tilde D_{{\rm L} k\omega}^0
		+
		1
	\right).
\end{eqnarray} 
The correction to the drag is:
\begin{eqnarray}
\delta r
=
\frac{
	g^2_{\rm C} \tilde g' {\cal K}^2
     }
     {
	8 \pi^2 \tilde v_{\rm F}' \tilde v_{\rm F}^2 T
     }
\int
\frac{
	dk d\omega
	\left(
		4\pi \tilde v_{\rm F}
		{\rm Re}\/ \tilde D_{{\rm L}k\omega}^0 
		+
		1
	\right)
     }
     {
	\left[
		\cosh(\frac{
				\tilde v_{\rm F} \delta \omega_{\rm R}
			   }
		   	{
				2 \tilde v_{\rm F}' k T
		   	}
	     	     )
		+
		\cosh(\frac{
				\tilde v_{\rm F} k
		   	   }
		   	   {
				2T 
		   	   }
	     )
	\right]^2
     }.
\label{delta_r}
\end{eqnarray} 
Since the denominator in this integral is very small everywhere except near 
$\omega = \tilde v_{\rm F} k$,
we need to know the behavior of 
${\rm Re}\/ \tilde D_{\rm L}^0$
in this region. It is proven in Appendix~\ref{app_ReD} that:
\begin{eqnarray}
\label{ReD3}
4 \pi \tilde v_{\rm F} {\rm Re}\/ \tilde D_{{\rm L}k\omega}^0 + 1
=
\frac{\delta \omega_{\rm R}}{2 \tilde v_{\rm F} k}
- \frac{1}{12} \left(
			\frac{\tilde v_{\rm F}' k}{\tilde v_{\rm F}}
	  	\right)^2
\\
\nonumber
- \frac{1}{4} \left(
			\frac{\delta \omega_{\rm R}}{\tilde v_{\rm F} k}
		\right)^2
-\frac{\pi^2}{24}
\frac{T^2}{\tilde \epsilon_{\rm F}^2}
+ O(T^3),
\end{eqnarray} 
where it is assumed that
$\tilde v_{\rm F} k = O(T)$,
and 
$\tilde v_{\rm F} \delta \omega_{\rm R}/\tilde v_{\rm F}' k = O(T)$.

At $T=0$ the above expansion becomes:
\begin{eqnarray}
4 \pi \tilde v_{\rm F} {\rm Re}\/ \tilde D_{pk\omega}^0 + 1
=
-\frac{p \delta \omega_p}{ 2 \tilde v_{\rm F} k }
- \frac{(\delta \omega_p)^2}{ 4 \tilde v_{\rm F}^2 k^2 }
- \frac{(\tilde v_{\rm F}' k )^2}{12 \tilde v_{\rm F}^2 }.
\end{eqnarray} 
This expansion is derived in \cite{drag_PRB,drag_arXiv}, see Eq.~(51) of
these references. 

Finally, placing Eq.~(\ref{ReD3}) into Eq. (\ref{delta_r}) one finds 
\begin{eqnarray}
\delta r \propto T^4.
\label{delta_r_fin}
\end{eqnarray}
This result may be obtained without performing actual integration. It is
enough to use the dimensional analysis:
\begin{eqnarray}
4 \pi \tilde v_{\rm F} {\rm Re}\/ \tilde D_{pk\omega}^0 + 1
= O(T^2),
\\
dk = O(T),
\\
d\omega = O(\delta \omega_{\rm R}) = O(T^2).
\end{eqnarray}
From these Eq.~(\ref{delta_r_fin}) follows.

Consequently, at low temperature the interaction correction $\delta r$ to
the Coulomb drag resistivity $r$ vanishes quicker than $T^2$. Thus, the
non-interacting quasiparticle result, Eq.(\ref{r_free}), suffices to capture
the leading behavior of $r$ at
$T\rightarrow 0$.

\section{Conclusions}
\label{conclusions}

Using bosonization-fermionization trick we mapped the Tomonaga-Luttinger
Hamiltonian with non-linear dispersion on the Hamiltonian of the
quasiparticles with irrelevant interaction. This mapping allows us to
evaluate the density-density propagator of the Tomonaga-Luttinger model
with non-linear dispersion. The propagator itself was used to calculate the
temperature dependence of the Coulomb drag resistivity $r$. It was
established that $r \propto T^2$ at low $T$ for both interacting and free
fermions. The irrelevant quasiparticle interaction introduces additional
correction which vanishes as $T^4$.

\section{Acknowledgements}

The author is grateful to Dr. D.Aristov who pointed out the mistake in
\cite{drag_PRB,drag_arXiv}. Support is provided by the RFBR grants
No.~08-02-00212, and 09-02-00248.

\appendix

\section{Some details of $\tilde B^0_{pk\omega}$ calculation}
\label{app_B0}

Here we fill the gap between Eq.~(\ref{B_qpp}) and Eq.~(\ref{B_finT}).
Observe that:
\begin{eqnarray}
n_{pq-k/2} - n_{pq+k/2}
=
\frac{
	\sinh \left(
			\frac{
				\tilde \varepsilon_{pq+k/2}
				-
				\tilde \varepsilon_{pq-k/2}
			     }
			     {
				2T
			     }
	     \right)
     }
     {
	2 \cosh \left(
			\frac{
				\tilde \varepsilon_{pq-k/2}
			     }
			     {
				2T
			     }
		\right)
	 \cosh \left(
			\frac{
				\tilde \varepsilon_{pq+k/2}
			     }
			     {
				2T
			     }
		\right)
     }.
\end{eqnarray}
The delta-function in Eq.~(\ref{B_qpp}) enforces 
$
\tilde \varepsilon_{pq+k/2} - \tilde \varepsilon_{pq-k/2} = \omega.
$
Therefore:
\begin{eqnarray} 
\label{B_integral}
\tilde B_{pk\omega}^0
=
\int dq
\frac{
	\sinh\left(
		\frac{\omega}{2T}
     	\right)
	\delta(\omega - p \tilde v_{\rm F} k - 2 \tilde v_{\rm F}' k q)
     }
     {
	2 \cosh(\frac{\tilde \varepsilon_{pq-k/2}}{2T})
          \cosh(\frac{\tilde \varepsilon_{pq+k/2}}{2T})
     }
\\
\nonumber
=
\frac{
	\sinh\left(
		\frac{\omega}{2T}
	     \right)
     }
     {
	2 |\tilde v_{\rm F}' k |
     }
\int dq
\frac{
	\delta(\frac{\omega - p \tilde v_{\rm F} k}
		    {2 \tilde v_{\rm F}' k} - q)
     }
     {
	2 \cosh(\frac{\tilde \varepsilon_{pq-k/2}}{2T})
          \cosh(\frac{\tilde \varepsilon_{pq+k/2}}{2T})
     }.
\end{eqnarray}
Substituting 
$q=(\omega - p \tilde v_{\rm F} k)/{2 \tilde v_{\rm F}' k}$,
as required by the delta-function, into expressions for
$\tilde \varepsilon_{pq + k/2}$
we obtain:
\begin{eqnarray}
q+k/2 = \frac{
		\delta \omega_p + \tilde v_{\rm F}' k^2
	     }
	     {
		2 \tilde v_{\rm F}' k 
	     },
\label{q+k/2}
\\
\varepsilon_{p q + k/2} 
=
\frac{
	\left(
		\delta \omega_p + \tilde v_{\rm F}' k^2
	\right)
	\left(
		\delta \omega_p + 2p \tilde v_{\rm F} k + \tilde v_{\rm F}' k^2
	\right)
     }
     {
	4 \tilde v_{\rm F}' k^2
     },
\label{energyI}
\\
\delta \omega_p = \omega - p \tilde v_{\rm F} k.
\end{eqnarray}
One can check that the right-hand side of Eq.~(\ref{energyI}) vanishes both at
$\delta \omega_p = - \tilde v_{\rm F}' k^2$
and at
$\delta \omega_p = 2 \tilde v_{\rm F} k - \tilde v_{\rm F}' k^2$. Of these
two roots only the first is physical, the second is spurious. The spurious
root appears due to unjustified extrapolation of Eq.~(\ref{dispersion}) to
large momenta 
$\sim \tilde v_{\rm F} / \tilde v_{\rm F}'$.
Indeed, using
Eq.~(\ref{q+k/2}), we find that the momentum $q+k/2$ is close to zero for
the physical root and close to
$-p\tilde v_{\rm F}/\tilde v_{\rm F}'$
for spurious root. Since our treatment is valid near the Fermi points only
we replace:
\begin{eqnarray}
\delta \omega_p + 2p \tilde v_{\rm F} k + \tilde v_{\rm F}' k^2
\approx
 2 p \tilde v_{\rm F} k.
\end{eqnarray}
In the above equation we neglected the terms, which are quadratic in $T$,
see Eq.~(\ref{region_kw}). Consequently:
\begin{eqnarray}
\tilde \varepsilon_{pq+k/2}
\approx
\frac{
	\tilde v_{\rm F}
     }
     {
	2 p \tilde v_{\rm F}' k
     }
(\delta \omega_p + \tilde v_{\rm F}' k^2).
\label{energyII}
\end{eqnarray}
With the help of the same arguments one derives:
\begin{eqnarray}
\tilde \varepsilon_{pq-k/2}
\approx
\frac{
	\tilde v_{\rm F}
     }
     {
	2 p \tilde v_{\rm F}' k
     }
(\delta \omega_p - \tilde v_{\rm F}' k^2).
\label{energyIII}
\end{eqnarray}
The expressions Eq.~(\ref{energyII}) and Eq.~(\ref{energyIII}) allow us to
perform integration over $q$ in Eq.~(\ref{B_integral}):
\begin{eqnarray} 
\tilde B_{pk\omega}^0
=
\frac{
	\sinh\left(
		\frac{\omega}{2T}
	     \right)
     }
     {
	4 |\tilde v_{\rm F}' k |
	\cosh(\tilde v_{\rm F} 
	      \frac{
			\delta \omega_p - \tilde v_{\rm F}' k^2
		   }
		   {
			4 \tilde v_{\rm F}' k T
		   }
	     )
	\cosh(\tilde v_{\rm F}
	      \frac{
			\delta \omega_p + \tilde v_{\rm F}' k^2
		   }
		   {
			4 \tilde v_{\rm F}' k T
		   }
	     )
     }.
\end{eqnarray}
This is the quasiparticle spectral function for $T>0$.

\section{Calculation of $I(T)$}
\label{app_I(T)}

The integral $I(T)$ is defined as:
\begin{eqnarray}
\label{r^0A}
I(T)
=
\int_0^{+\infty} dk
\int_{0}^{+\infty} d\omega
\times
\\
\nonumber
\frac{
		1
     }
     {
		\cosh^{2}(\tilde v_{\rm F} 
			      \frac{
				     \omega - \tilde v_{\rm F} k 
					- \tilde v_{\rm F}' k^2
		   		   }
		   		   {
					4 \tilde v_{\rm F}' k T
				   }
	     		     )
		\cosh^{2}(\tilde v_{\rm F}
	     		      \frac{
				     \omega - \tilde v_{\rm F} k 
					+ \tilde v_{\rm F}' k^2
		                   }
		                   {
					4 \tilde v_{\rm F}' k T
		                   }
	                     )
     }.
\end{eqnarray}
In the above integral we introduce the new variable:
\begin{eqnarray}
\Omega =
\frac{
	\tilde v_{\rm F}
     }
     {
	4|\tilde v_{\rm F}'| k T
     }
(\omega - \tilde v_{\rm F} k - |\tilde v_{\rm F}'| k^2),
\\
I
=
\frac{
	4 |\tilde v_{\rm F}'| T
     }
     {
	\tilde v_{\rm F}
     }
\int_0^{+\infty} dk
\int_{-\Omega_0}^{+\infty} 
   \frac{
		k d \Omega
	}
	{
		\cosh^2 \Omega \cosh^2 (\Omega 
					+ 
					\frac{
						\tilde v_{\rm F} k
					     }
					     {
						2T
					     }
					)
        },
\\
\Omega_0 = \frac{
			\tilde v_{\rm F}^2
		}
		{
			4|\tilde v_{\rm F}'|  T
		}
	+ \frac{
			\tilde v_{\rm F} k
		}
		{
			4 T
		}.
\end{eqnarray}
If temperature is low:
$T \ll \tilde \epsilon_F$,
where 
\begin{eqnarray} 
\tilde \epsilon_{\rm F}
=
\frac{\tilde v_{\rm F}^2}{ 4 \tilde v_{\rm F}'},
\end{eqnarray} 
then $\Omega_0 \gg 1$.
In such a situation the integrand is exponentially small for 
$\Omega < -\Omega_0$:
\begin{eqnarray}
\cosh^{-2} \Omega \cosh^{-2} (\Omega 
			+ 
			\tilde v_{\rm F} k/2T
			)
\le
\cosh^{-2} \Omega 
\\
\nonumber
< 4 \exp(-2\Omega_0) \ll 1.
\end{eqnarray}
Therefore, we may extend the integration interval:
\begin{eqnarray}
\int_{-\Omega_0}^{+\infty} d\Omega \ldots 
\approx
\int_{-\infty}^{+\infty} d\Omega \ldots.
\end{eqnarray}
The integral $I$ can be expressed as:
\begin{eqnarray}
I =
\frac{
	4 |\tilde v_{\rm F}'|  T
     }
     {
	\tilde v_{\rm F}
     }
\int_0^{+\infty} kdk
\int_{-\infty}^{+\infty} 
	\frac{
		d\Omega
	     }
	     {
		\cosh^{2}\Omega
		\cosh^{2}(\Omega + \frac{\tilde v_{\rm F} k}{2T})
	     }
\\
\nonumber
=
\frac{
	 16 |\tilde v_{\rm F}'|  T^3
     }
     {
	\tilde v_{\rm F}^3
     }
\int_{-\infty}^{+\infty} 
	\frac{
		d\Omega
	     }
	     {
		\cosh^{2}\Omega 
	     }
\int_0^{+\infty}
	\frac{
		QdQ
	     }
	     { 
		\cosh^{2}(\Omega + Q)
	     }.
\end{eqnarray} 
Integrating over $Q$ by parts we derive:
\begin{eqnarray}
I(T) =
\frac{
	 16 |\tilde v_{\rm F}'|  T^3
     }
     {
	\tilde v_{\rm F}^3
     }
\int_{-\infty}^{+\infty} 
	\frac{
		d\Omega
	     }
	     {
		\cosh^{2}\Omega 
	     }
	f(\Omega),
\\
f(\Omega) = \int_0^{+\infty}
		\left[
			1 - \tanh (\Omega + Q)
		\right]
		dQ.
\end{eqnarray} 	
Since
$f(\Omega) \rightarrow 0$
when
$\Omega \rightarrow +\infty$,
we can integrate by parts over $\Omega$ as follows:
\begin{eqnarray}
I(T) =
-\frac{
	 16 |\tilde v_{\rm F}'|  T^3
     }
     {
	\tilde v_{\rm F}^3
     }
\int_{-\infty}^{+\infty}
(\tanh \Omega + 1) f'(\Omega) d \Omega,
\\
f'(\Omega) = - \int_0^{+\infty} \frac{d\Omega}{\cosh^2 (\Omega + Q)} 
= \tanh \Omega - 1.
\end{eqnarray} 
Therefore:
\begin{eqnarray}
I(T)
=
\frac{
	 16 |\tilde v_{\rm F}'|  T^3
     }
     {
	\tilde v_{\rm F}^3
     }
\int_{-\infty}^{+\infty}
(1 - \tanh^2 \Omega)  d \Omega.
\end{eqnarray}
The last integral can be calculated easily, and $I(T)$ is:
\begin{eqnarray}
I(T)
=
\frac{
	 32 |\tilde v_{\rm F}'|  T^3
     }
     {
	\tilde v_{\rm F}^3
     }.
\end{eqnarray}
This concludes our derivation of $I(T)$.

\section{Expansion of ${\rm Re}\/ \tilde D_{\rm L}$}
\label{app_ReD}

In this Appendix we evaluate 
${\rm Re}\/ \tilde D_{{\rm L}k\omega}^0$
when
$\omega \approx \tilde v_{\rm F} k$.
We start with the following expression:
\begin{eqnarray}
{\rm Re}\/ \tilde D_{{\rm L}k\omega}^0
=
{\rm Re}\/ \int_{q} \frac{
                n_{{\rm L}q-k/2} - n_{{\rm L}q+k/2} 
              }
              {
                \omega + \tilde \varepsilon_{{\rm L}q-k/2} 
                - \tilde \varepsilon_{{\rm L}q+k/2} + i0
              }.
\end{eqnarray} 
From which one gets:
\begin{eqnarray}
{\rm Re}\/ \tilde D_{{\rm L}k\omega}^0
=
-\frac{1}{2 \tilde v_{\rm F}' k} 
{\rm Re}\/ \int_{q} \frac{
                n_{{\rm L}q-k/2} - n_{{\rm L}q+k/2} 
              }
              {
                q - \delta \omega_{\rm L}/2 \tilde v_{\rm F}' k + i0
              }.
\end{eqnarray} 
In the last expression we integrate by parts to obtain:
\begin{eqnarray}
{\rm Re}\/ \tilde D_{{\rm L}k\omega}^0
=
\frac{1}{2 \tilde v_{\rm F}' k}
\int_q \ln \left|
		\frac{
			q + \frac{k}{2}
			- 
			\frac{\delta \omega_{\rm L}}{ 2 \tilde v_{\rm F}' k}
		     }
		     {
			q - \frac{k}{2}
			- 
			\frac{\delta \omega_{\rm L}}{ 2 \tilde v_{\rm F}' k}
		     }
	   \right|
	\frac{
		dn_{{\rm L}q}
	     }
	     {
		dq
	     }.
\end{eqnarray} 
Since we want to derive 
${\rm Re}\/ \tilde D_{{\rm L}k\omega}^0$ 
near the point
$\delta \omega_{\rm R} \approx 0$,
it is convenient to substitute 
$\delta \omega_{\rm L}$
with the equal quantity
$\delta \omega_{\rm R} + 2 \tilde v_{\rm F} k$.
After such substitution and some transformations the expression for
${\rm Re}\/ \tilde D_{{\rm L}k\omega}^0$
becomes:
\begin{eqnarray}
{\rm Re}\/ \tilde D_{{\rm L}k\omega}^0
=
\frac{1}{2 \tilde v_{\rm F}' k}
\int_q \ln \left|
		\frac{
			1 - \frac{\tilde v_{\rm F}' q }{\tilde v_{\rm F}}
			-
			\frac{\tilde v_{\rm F}' k}{2 \tilde v_{\rm F}}
			+ 
			\frac{\delta \omega_{\rm R}}{ 2 \tilde v_{\rm F} k}
		     }
		     {
			1 - \frac{\tilde v_{\rm F}' q }{\tilde v_{\rm F}}
			+
			\frac{\tilde v_{\rm F}' k}{2 \tilde v_{\rm F}}
			+ 
			\frac{\delta \omega_{\rm R}}{ 2 \tilde v_{\rm F} k}
		     }
	   \right|
	\frac{
		dn_{{\rm L}q}
	     }
	     {
		dq
	     }.
\end{eqnarray} 
Logarithm in this formula may be expanded in orders of small parameter 
$T/\tilde \epsilon_{\rm F}$.
Indeed, it is true that
$\tilde v_{\rm F} |k| = O(T)$
and
$\tilde v_{\rm F} |\delta \omega_{\rm R}/ \tilde v_{\rm F}' k| = O(T)$.
Observe also:
\begin{eqnarray}
\int q dq 
\frac{dn_{{\rm L}q}}{dq}
=
\int d \epsilon
\left(
	\frac{
		\epsilon
	      }
	      {
		\tilde v_{\rm F}
	      }
	-\frac{
		\tilde v_{\rm F}' \epsilon^2
	      }
	      {
		\tilde v_{\rm F}^3
	      }
\right)
\frac{d}{d\epsilon}
\frac{1}{1+\exp (\epsilon/T)}
\\
\nonumber
=
\frac{
	\tilde v_{\rm F}'
     }
     {
	4 \tilde v_{\rm F}^3 T
     }
\int d \epsilon
\frac{
	\epsilon^2
     }
     {
	\cosh^2 (\epsilon/2T)
     }
=
\frac{
	\pi^2  \tilde v_{\rm F}' T^2
     }
     {
	3 \tilde v_{\rm F}^3
     },
\end{eqnarray} 
where we used 
$q \approx - \tilde \varepsilon_{{\rm L}q}/\tilde v_{\rm F} 
+ \tilde v_{\rm F}' \tilde \varepsilon_{{\rm L}q}^2/\tilde v_{\rm F}^3$
and the identity:
\begin{eqnarray}
\int_{-\infty}^{+\infty} 
\frac{
	x^2 dx 
     }
     {
	\cosh^2 x
     }
=
\frac{
	\pi^2
     }
     {
	6
     }.
\end{eqnarray} 
Using similar procedure one proves:
\begin{eqnarray}
\int q^2 dq 
\frac{dn_{{\rm L}q}}{dq}
=
\frac{
	\pi^2 T^2
     }
     {
	3 \tilde v_{\rm F}^2
     }.
\end{eqnarray} 
Therefore, both $q$ and $q^2$, upon integration, contribute terms of order
$T^2$. Thus:
\begin{eqnarray}
{\rm Re}\/ \tilde D_{{\rm L}k\omega}^0
=
-\frac{1}{2 \tilde v_{\rm F}}
\int_q 
	\frac{
		dn_{{\rm L}q}
	     }
	     {
		dq
	     }
\left[
	1
	+
	\frac{ \tilde v_{\rm F}' q}{\tilde v_{\rm F}}
	-
	\frac{\delta \omega_{\rm R}}{2 \tilde v_{\rm F} k}
\right.
\label{ReD}
\\
\nonumber
\left.
	+
	\frac{1}{12}
	\left(
		\frac{
			\tilde v_{\rm F}' k
		     }
		     {
			\tilde v_{\rm F} 
		     }
	\right)^2
	+
	\left(
		\frac{
			\tilde v_{\rm F}' q
		     }
		     {
			\tilde v_{\rm F}
		     }
	\right)^2
	+
	\frac{1}{4}
	\left(
		\frac{
			\delta \omega_{\rm R}
		     }
		     {
			\tilde v_{\rm F} k
		     }
	\right)^2
\right] + O(T^3).
\end{eqnarray} 
Since 
$\int_q dn_{{\rm L} q} / dq = 1/(2\pi)$,
Eq.~(\ref{ReD}) may be written as:
\begin{eqnarray}
4 \pi \tilde v_{\rm F} {\rm Re}\/ \tilde D_{{\rm L}k\omega}^0 + 1
=
\frac{\delta \omega_{\rm R}}{2 \tilde v_{\rm F} k}
- \frac{1}{12} \left(
			\frac{\tilde v_{\rm F}' k}{\tilde v_{\rm F}}
	  	\right)^2
\\
\nonumber
- \frac{1}{4} \left(
			\frac{\delta \omega_{\rm R}}{\tilde v_{\rm F} k}
		\right)^2
-\frac{\pi^2}{24}
\left(
	\frac{T}{\tilde \epsilon_{\rm F}}
\right)^2
+ O(T^3).
\end{eqnarray}
This is the expansion we need.


\end{document}